\documentclass[aps,prl,amsmath,amssymb,twocolumn,longbibliography, superscriptaddress]{revtex4-2}
\usepackage{graphicx}
\usepackage{dcolumn}
\usepackage{bm}
\usepackage{mathrsfs}
\usepackage{subfigure}
\usepackage{natbib}
\usepackage{amssymb}
\usepackage{multirow}
\usepackage{float}
\usepackage{color}
\usepackage{ulem}
\usepackage{mathrsfs}
\usepackage[colorlinks,linkcolor=blue,anchorcolor=blue,citecolor=blue,urlcolor=blue]{hyperref}

\begin{document}

\title{Emergence of Surface Superconductivity through Interference in Superconducting-proximity Topological Insulators}


\author{Yajiang Chen}
\email{yjchen@zstu.edu.cn}
\affiliation{Key Laboratory of Optical Field Manipulation of Zhejiang Province, Department of Physics, Zhejiang Sci-Tech University, 310018 Zhejiang, China}

\author{Ke-Ji Chen}
\affiliation{Key Laboratory of Optical Field Manipulation of Zhejiang Province, Department of Physics, Zhejiang Sci-Tech University, 310018 Zhejiang, China}

\author{Jia-Ji Zhu}
\affiliation{School of Science, Chongqing University of Posts and Telecommunications, Chongqing 400065, China}

\author{A. A. Shanenko}
\affiliation{HSE University, 101000 Moscow, Russia}

\date{\today}
\begin{abstract}
Superconducting-proximity topological insulators (STIs) have garnered significant research attention over the past two decades. In this Letter, we demonstrate that a low-dimensional STI in the topological-nontrivial phase (TP) exhibits an interference-induced surface (boundary) superconductivity with the surface critical temperature $T_{cs}$ significantly higher than the bulk one $T_{cb}$. Such a surface superconductivity is built due to the interference of the scattering quasiparticle states, rather than due to the presence of the topological bound states (TBSs). As the system goes deeper into the TP, the surface exhibits a crossover from the interference- to TBS-induced phase, where the surface enhancement of superconductivity is governed by the TBSs. Our study unveils a substantial variation in the maximal $T_{cs}$ along this crossover, attaining values being twice the maximal bulk critical temperature of the STI. Beyond shedding light on the nature of surface superconductivity in STIs, our study introduces a tangible method for experimentally manipulating their critical superconducting temperatures.
\end{abstract}
\maketitle

\textit{Introduction.—}Superconducting-proximity topological insulators (STIs) have been studied intensively in the past two decades, due to exotic behavior of the decoherence-immune topological bound states (TBSs) under the superconducting interactions~\cite{Fu2008, Bernevig2013, Shen2017, Chen2022a,Li2023,Flensberg2021}. The point of common interest in the context of the STIs is the Majorana quasiparticles considered as a solid candidate for qubits~\cite{Kitaev2001}. It has been predicted that these Majorana states can appear in the core of an $s-$wave superconducting vortex~\cite{Fu2008, Chen2022a,Li2023,Xu2015}, at the nodes of unconventional order parameters~\cite{Linder2010, Roy2010} or at the ferromagnet topological insulator/superconductor interface~\cite{Tanaka2009, Beenakker2013}. Moreover, besides the Majorana states, the STIs exhibit an unconventional Josephson effect~\cite{Williams2012}, odd-frequency superconductivity (SC)~\cite{Krieger2020}, and coexistence of the TBSs and SC in bismuth and stanene ultrathin films~\cite{Sun2017,Zhao2022}, etc. 

The superconducting properties of the STIs have been investigated in a number of works~\cite{Stanescu2010, Lababidi2011, Black-Schaffer2011, Mizushima2014, Park2020, Liu2022,Zhu2023}. In particular, the suppression of the pair potential $\Delta({\bf r})$ at the STI interface between a superconductor and topological insulator has been reported~\cite{Lababidi2011,Black-Schaffer2011,Mizushima2014,Park2020} and verified by the observed Andreev spectra~\cite{Zareapour2012}. Such a suppression assumes the corresponding drop in the superconducting critical temperature $T_c$. However, a study based on the dynamical mean-field theory has shown that a 2D attractive ($s$-wave) Hubbard model with Rashba spin-orbital coupling and a Zeeman field exhibits an enhanced $T_c$~\cite{Nagai2016}. 
Further research and clarifications are warranted to reconcile these contradictory findings and advance our understanding, paving the way for controllable manipulations of the superconducting properties of STIs.
 
Recently, an exotic interference-induced surface SC in the absence of magnetic fields has been predicted in the $s$-wave superconducting Hubbard model~\cite{Croitoru2020, Samoilenka2020,Samoilenka2020a}, where the surface pair potential survives at the temperatures between the bulk $T_{cb}$ and surface $T_{cs}$ critical temperatures ($T_{cs}>T_{cb}$) due to the interference of non-localized (scattering) quasiparticles. The enhancement $\tau=|T_{cs}-T_{cb}|/T_{cb}$ of the surface SC goes up to about $70\%$ by adjusting the Debye energy~\cite{Bai2023} and can be tailored by an applied electric field~\cite{Bai2023b}. Therefore, the two questions arise - can interference-induced SC coexist with the TBSs in the STIs, and, if yes, what impact does this coexistence exert on $T_{cs}$ and $\tau$? This holds particular significance for low-dimensional STIs because of the pronounced proximity effects. 

In this Letter, we demonstrate that a low-dimensional STI can exhibit a robust interference-induced surface SC alongside the TBSs. By changing the microscopic parameters of the model, the interference-induced surface SC can be transformed into the TBS-induced surface SC. In the interference-induced phase, the TBS contributes less than $1\%$ to the surface SC, while the non-localized scattering quasiparticles yield a contribution of about $99\%$. Conversely, this situation is reversed in the TBS-induced surface phase. Moreover, our results reveal that the maximum of $T_{cs}$ in the interference- and TBS-induced boundary phases surpasses the maximum of $T_{cb}$~(far beyond the boundaries) by factors of $1.9$ and $2.5$, respectively.

\begin{figure*}[ht]
\centering
\includegraphics[width=0.95\linewidth]{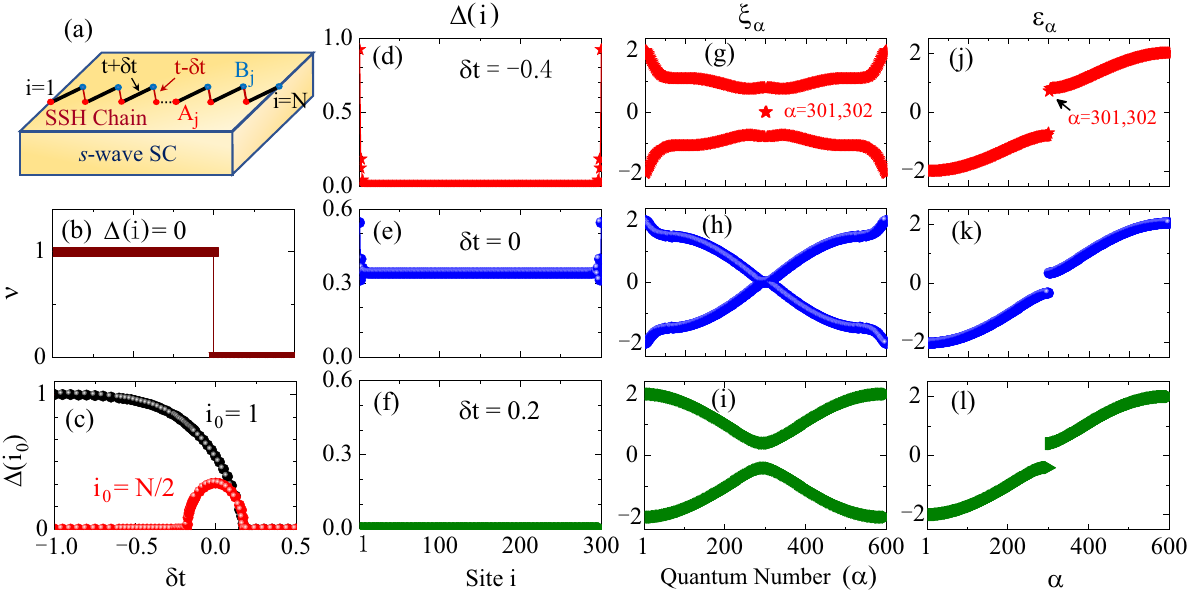}
\caption{
(a) A superconducting-proximity SSH chain. (b) The topological invariant $\nu$ versus $\delta t$ for the SSH chain without superconducting correlations. (c) The surface $\Delta_s=\Delta(i_0=1)$ and the bulk $\Delta_b=\Delta(i_0=N/2)$ pair potentials as functions of $\delta t$. (d-l) The spatial pair potential $\Delta(i)$ together with the single-electron and quasiparticle energies $\xi_\alpha$ and $\epsilon_\alpha$ for $\delta t=-0.4$, $0$ and $0.2$ at $T=0$.}
\label{fig1}
\end{figure*}

\textit{Model.—}For illustration, we take a one-dimensional Su-Schrieffer-Heeger (SSH) chain at the half-filling level as the topological insulator~\cite{Bernevig2013, Shen2017, Su1979, Heeger1988, Guo2011,Wang2023,Xing2023} deposited on top of an $s-$wave superconductor, see Fig.~\ref{fig1}(a). Thus, the model Hamiltonian includes the standard SSH contribution~\cite{Bernevig2013,Shen2017} $H_{0} = -\sum_{j\sigma} (t+\delta t)  c^{\dagger}_{A_j\sigma}c_{B_j\sigma}-\sum_{j\sigma} (t-\delta t)  c^{\dagger}_{A_{j+1}\sigma}c_{B_j\sigma} - \sum_{j\sigma}\mu \big(c^{\dagger}_{A_j\sigma}c_{A_j\sigma} + c^{\dagger}_{B_j\sigma}c_{B_j\sigma}\big)$ together with the effective superconducting-proximity interaction $H_{SC} = -g \sum_i n_{i\uparrow}n_{i\downarrow}$. Here, indexes $i(=1,...,N)$ and $j$ enumerate the chain and sublattice $A(B)$ sites, an odd (even) $i$ corresponds to the $A(B)$ sublattice. $c_{A(B)_j\sigma}$ and $c^{\dagger}_{A(B)_j\sigma}$ are the annihilation and creation operators of an electron with the spin $\sigma$ on the sites $A(B)_j$, $n_{i\sigma}$ is the site-dependent electron number operator. $t\pm\delta t$ are the hopping amplitudes in a unit cell and between the neighboring unit cells, $\mu$ is the chemical potential, and $g(>0)$ is the attractive interaction strength. Notice that our consideration follows the phenomenological Fu-Kane model~\cite{Fu2008} that takes into account the tunneling of Cooper pairs into the topological insulator by including the $s$-wave pairing term $H_{SC}$ in the Hamiltonian. 

The spatial distribution of the superconducting condensate can be derived from the Bogoliubov-de Gennes (BdG) equations. The BdG are obtained by applying the mean-field approximation $H_{SC}' = \sum_{j}\big[ \Delta(A_j)c^{\dagger}_{A_j\uparrow}c^{\dagger}_{A_j\downarrow} + \Delta(B_j) c^{\dagger}_{B_j\uparrow} c^{\dagger}_{B_j\downarrow} + \rm{H.c.} \big]$ and diagonalizing the mean-field Hamiltonian $H_{\rm eff} = H_{0} + H_{SC}'$~\cite{Gennes1966}. The corresponding BdG equations read
\begin{subequations}\label{bdg}
\begin{align}
\epsilon_\alpha u_\alpha(i) & =  \sum_{i'} H_{ii'}u_\alpha(i') + \Delta(i) v_\alpha(i) \\
\epsilon_\alpha v_\alpha(i) & =  \Delta^*(i)u_\alpha(i)  - \sum_{i'} H^*_{ii'} v_\alpha(i'),
\end{align}
\end{subequations}
where $H_{ii'} = -\sum_{\eta=\pm1}[t-(-1)^i{\rm sgn}(\eta)\delta t] \delta_{i',i+\eta}-\mu\delta_{ii'}$, and $\{\epsilon_\alpha, \,u_\alpha(i), \,v_\alpha(i)\}$ are the quasiparticle energies and wavefunctions, with $\alpha$ enumerating the quasiparticle states in the energy ascending manner. To investigate surface effects, the open boundary conditions are employed for $u_\alpha(i)$ and $v_\alpha(i)$. In addition, $\Delta(i)$ in Eqs.~(\ref{bdg}) is the pair potential that obeys the standard ($s$-wave) self-consistency relation
\begin{equation}\label{pair_potential}
\Delta(i)=g\langle c_{i\uparrow}c_{i\downarrow}\rangle=g\sum_{\epsilon_\alpha\geq0} u_\alpha(i) v_\alpha^*(i)[1-2f(\epsilon_\alpha)]
\end{equation}
with $f_\alpha=f(\epsilon_\alpha)$ being the Fermi-Dirac function.

The electron half-filling level, i.e., $\bar{n}_e = \sum_{i} n_e(i)/N=1$, with $n_e(i)$ the average site occupation (electron density), is guaranteed by the self-consistent calculation of the chemical potential $\mu$ according to $n_e(i) = 2\sum_{\alpha}\big[ f_\alpha |u_\alpha(i)|^2 + (1-f_\alpha) |v_\alpha(i)|^2\big]$.

Below, the energy-related quantities, such as $\Delta(i)$, $\delta t$, $\epsilon_\alpha$, and $g$, are given in the units of the hopping parameter $t$, while $T$ is in the units of $t/k_B$, with $k_B$ the Boltzmann constant. The following parameters are used in our calculations: $\bar{n}_e=1$, $g=2$, and $N=301$. The self-consistency convergence accuracy for $\Delta(i)$ is $10^{-12}$. 

\textit{Interference-induced surface SC.—}To begin the discussion of our numerical results, we consider the effects of varying the hopping staggering parameter $\delta t$ at $T=0$. The topological invariant $\nu$~(the winding numbers) is shown versus $\delta t$ in Fig.~\ref{fig1}(b) for $\Delta(i)=0$. One can see that a topological phase transition occurs at $\delta t=0$ in the normal SSH model~\cite{Shen2017} at the half filling so that the system is in the topological non-trivial phase (TP) for negative $\delta t$ and in the topological trivial phase (TTP) for positive $\delta t$. The presence of the TP and TTP phases is reflected in the crucial dependence of the pair potential on $\delta t$, see Figs.~\ref{fig1}(b) and (c), where the surface $\Delta_s=\Delta(i=1)=\Delta(i=N)$ and bulk $\Delta_b=\Delta(i=N/2)$ pair potentials are shown versus $\delta t$ at $T=0$. One can see that the superconducting correlations disappear when the system goes deep in the TTP while they are nonzero (both at the surface and in bulk) deep in the TP.

Usually, it is believed that only the surface of an STI is superconducting, so one could expect that the pair potential in the superconducting-proximity SSH chain is nonzero only near the chain edges. However, surprisingly, we find that $\Delta_b$ can be tuned by $\delta t$, as seen from Fig.~\ref{fig1}(c). The curve of $\Delta_b$ exhibits a dome structure around the point $\delta t=0$: $\Delta_b=0$ for $|\delta t| \geq \delta t_c=0.18$, and the maximum is reached at $\delta t=0$. Then, the superconducting correlations smear the topological phase transition and lead to the formation of the dome structure in the dependence of $\Delta_b$ on $\delta t$. In turn, the surface pair potential $\Delta_s$ can also be controlled by $\delta t$. It decreases monotonically with increasing $\delta t$, vanishing above $\delta t=0.18$. 

Further information about the dependence of our results on $\delta t$ can be obtained from Figs.~\ref{fig1}(d-l). In particular, in Figs.~\ref{fig1}(d-f) one can see the site-dependent pair potential $\Delta(i)$ calculated at $T=0$ for $\delta t=-0.4$, $0$, and $0.2$, respectively. In Figs.~\ref{fig1}(g-i) we can find the single-particle energy~\cite{Shanenko2008, Chen2012} $\xi_\alpha = \sum_{ii'} u^*_\alpha(i') H_{ii'}u_\alpha(i)+v^*_\alpha(i')H_{ii'}v_\alpha(i)$ calculated for the same values of $\delta t$ and shown versus the quantum number $\alpha$. Finally, the corresponding quasiparticle energies $\epsilon_\alpha$ are given versus $\alpha$ in Figs.~\ref{fig1}(j-l).

Figures~\ref{fig1}(d, g, j) show the results corresponding to $\delta t=-0.4$. In this case, the system is deep in the TP regime, and we have quite significant surface pair potential $\Delta_s = 0.924$ while $\Delta_b$ vanishes. One can see that the single-particle spectrum $\xi_\alpha$ contains two branches and four zero-energy modes with $\alpha=299$, $300$, $301$, and $302$. Only the upper branch with the two zero modes $\alpha=301$ and $302$ is physical, see the selection rule in the self-consistency relation Eq.~(\ref{pair_potential}). When $\delta t$ increases up to $0$, as shown in Figs.~\ref{fig1}(e, h, k), the superconducting-proximity SSH model degenerates into the attractive Hubbard model~\cite{Chen2022, Bai2023, Bai2023b,Samoilenka2020}. Here, $\Delta_b$ goes up to $0.340$ while $\Delta_s$ drops to $0.544$. In this case the gap in the single-particle spectrum $\xi_\alpha$ is closed. Finally, for $\delta t = 0.2$, i.e. deep in the TTP, the site-dependent superconducting correlations disappear both in bulk and at the edges of the chain, and $\epsilon_\alpha$ approaches $\pm|\xi_\alpha|$. We find that the single-particle energy spectrum $\xi_\alpha$ in Figs.~\ref{fig1}(g, h, i) demonstrates clear signatures of the topological phase transition, including the opening of the gap and the appearance of the zero-energy modes. However, there are no zero-energy quasiparticle states in Figs.~\ref{fig1}(j, k, l), as the system is not a topological superconductor.

\begin{figure}[t]
\centering
\includegraphics[width=1\linewidth]{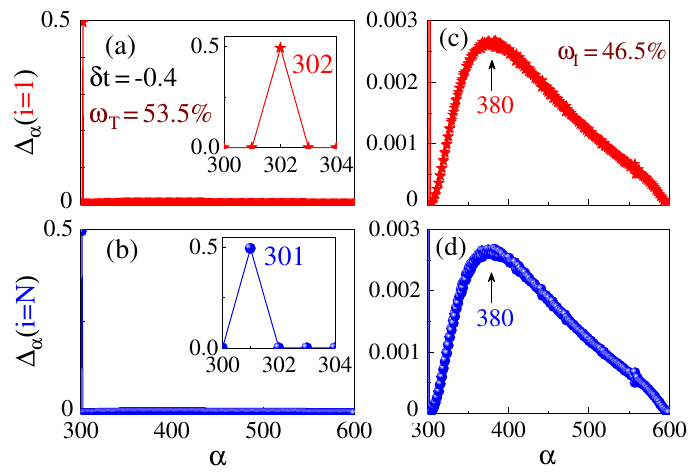}
\caption{(a, b) Single-species quasiparticle contributions $\Delta_\alpha(i=1,N)$ as functions of $\alpha$ for $\delta t=-0.4$ and $T=0$. The inserts and panels (c, d) are zoom-in views of (a, b).}
\label{fig2}
\end{figure}

Now, to go into more detail, we investigate what quasiparticle states are responsible for the enhanced surface pair potential $\Delta_s$ at both left ($i=1$) and right ($i=N$) sides of the system. By taking $\delta t = -0.4$ as an example, Figure~\ref{fig2} illustrates the single-species quasiparticle contributions $\Delta_\alpha(i=1,N)$ to $\Delta_s$ at $T=0$, utilizing the definition $\Delta_\alpha(i) =gu_\alpha(i)v^*_\alpha(i)[1-2f(\epsilon_\alpha)]$. Panels (c) and (d) are the zoom-in plots of Fig.~\ref{fig2}(a) and (d), respectively. One finds that the quasiparticle state with $\alpha=302$ gives the maximal contribution to the surface SC at $i=1$: $\Delta_{302}(i=1)=0.494$, which accounts for $\omega_T=53.5\%$ of $\Delta(i=1)$. The single-species contributions of all the other states are tiny, e.g., the second largest one comes from the state with $\alpha=380$, being about $0.29\%$. A similar picture takes place near the right edge. The only difference is that the maximal contribution comes from the state with $\alpha=301$, see Fig.~\ref{fig2}(b). 

\begin{figure}[t]
\centering
\includegraphics[width=1\linewidth]{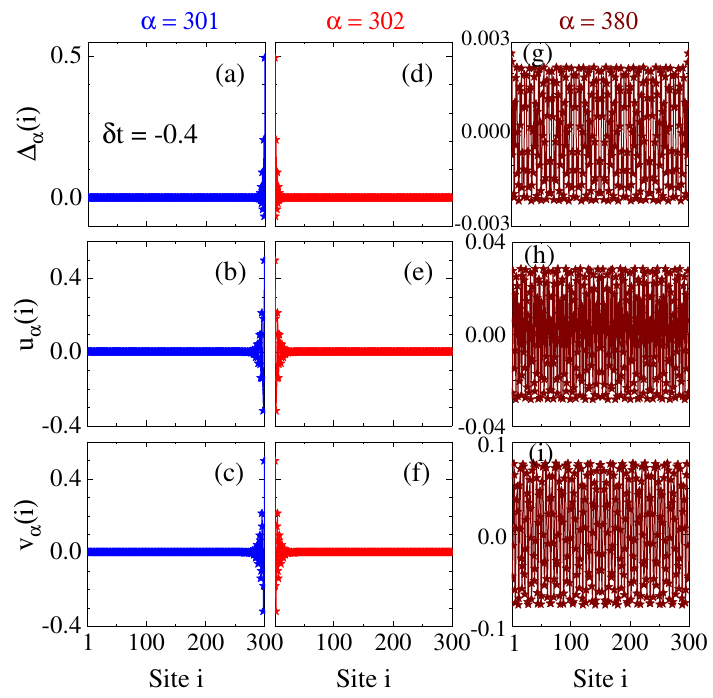}
\caption{(a-f) $\Delta_\alpha(i)$, $u_\alpha(i)$ and $v_\alpha(i)$ of the topological bound states (TBSs) with $\alpha=301$ and $302$ for $\delta t=-0.4$ at $T=0$. (g-i) The same but for the scattering quasiparticle state with $\alpha=380$.}
\label{fig3}
\end{figure}

It is important to remind that the states with $\alpha=301$ and $302$ for $\delta t=-0.4$ and $T=0$ have zero single-particle energy, as shown in Fig.~\ref{fig1}(g). One can find from Fig.~\ref{fig3} that the wavefunctions of these states are localized near the chain edges. Thus, the quasiparticles with $\alpha=301$ and $302$ are indeed related to the TBSs of the SSH model. Their inputs to the pair potential $\Delta_{\alpha=301,302}(i)$ are localized near the chain edges, see Fig.~\ref{fig3}(a, d), so that these states contribute only to the surface SC. The remaining quasiparticle states are not bound near the edges, being the scattering states. For example, one can see from Figs.~\ref{fig3}(g-i) that the quasiparticle wavefunctions and the single-species quasiparticle contribution to the pair potential for the states with $\alpha=380$ are spread all over the chain, with no sign of any localization. However, the constructive interference of all such scattering states produces a significant contribution to the surface pair potential [$\omega_I=1-\omega_T=46.5\%$, as seen in Fig.~\ref{fig2}(c)] while their destructive interference results in the suppression of $\Delta_b$, see the works~\cite{Chen2022, Bai2023,Bai2023b,Samoilenka2020}.

\begin{figure}[t]
\centering
\includegraphics[width=1\linewidth]{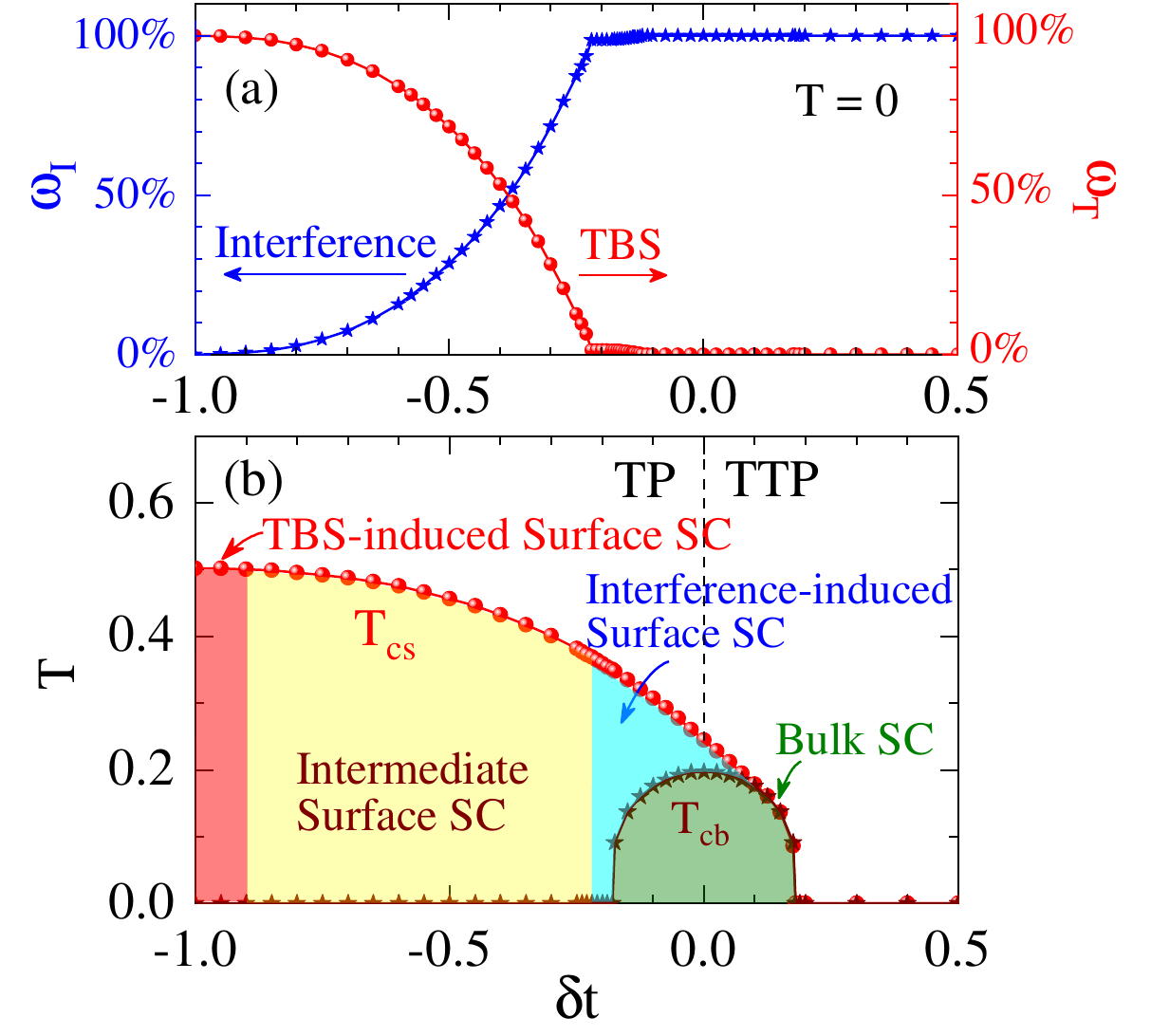}
\caption{(a) The contribution percentages of the TBSs and scattering states to $\Delta(i=1)$, i.e., $\omega_T$ and $\omega_I$, as functions of $\delta t$ at $T=0$. (b) The phase diagram of the superconducting-proximity SSH chain in the $T-\delta t$ plane. The curve with red solid spheres represents the surface critical temperature $T_{cs}$ versus $\delta t$, the curve with dark green stars corresponds to the $\delta t$-dependent bulk critical temperature $T_{cb}$. The dashed vertical line marks the boundary between the TP and TTP and correspond to $\delta t=0$.}
\label{fig4}
\end{figure}

The competition between the interference- and TBS-type contributions to the surface SC (i.e., $\omega_I$ versus $\omega_T$) is highly sensitive to $\delta t$. Figure~\ref{fig4}(a) shows $\omega_I$ and $\omega_T$ as functions of $\delta t$ for $g=2$ at $T=0$. One sees that both the TBS- and interference-type contributions vary significantly with changing $\delta t$. For example, $\omega_T$ drops from $100\%$ at $\delta t = -1$ to nearly $1\%$ at $\delta t = -0.22$ while $\omega_I$ increases from $0\%$ to almost $99\%$ in the same $\delta t$ range. Therefore, the interference-induced surface SC predominates for $\delta t \geq -0.22$. In the vicinity of $\delta t = -1$, the system shows the well-known TBS-type surface SC. The crossover intermediate regime, with the competing TBS- and interference-induced contributions, is realized for $-0.9<\delta t < -0.22$.

\textit{Tunable Critical Temperatures and Phase Diagram.—}Now, we study the upper (surface) and lower (bulk) critical temperatures of the superconducting-proximity SSH chain, i.e., $T_{cs}$ and $T_{cb}$ defined by the conditions $\Delta_{s}(T \geq T_{cs}) = 0$ and $\Delta_{b}(T \geq T_{cb}) = 0$. Figure~\ref{fig4}(b) shows $T_{cs}$ (red spheres) and $T_{cb}$ (wine-color stars) as functions of the hopping staggering parameter $\delta t$. The dependencies of $T_{cs}$ and $T_{cb}$ on $\delta t$ are similar to those of $\Delta_s$ and $\Delta_b$ at $T=0$ in Fig.~\ref{fig1}(c). This implies that the ratios $\gamma_{s,b}=\Delta_{s,b}(T=0)/k_BT_{cs,b}$ are not very sensitive to $\delta t$. In particular, we find that $\gamma_b\approx1.73$, which is close to the case of the conventional BCS superconductors $1.76$. In turn, $\gamma_s$ slightly increases from $2.00$ at $\delta t = -1$ to $2.23$ at $\delta t=0$. One can also learn from Fig.~\ref{fig4}(a) that $T_{cs}$ and $T_{cb}$ can be controlled and fine-tuned by changing $\delta t$. In particular, $T_{cs}$ is about $0.50$ at $\delta t = -1$, decreases with increasing $\delta t$, and vanishes for $\delta t \geq 0.18$. In turn, $T_{cb}$ is equal to $0$ for $|\delta t| \geq 0.18$ while reaching its maximum $0.20$ at $\delta t=0$. 

As seen in Fig.~\ref{fig4}(b), the $T_{cs}$ and $T_{cb}$ curves, together with the dashed vertical line $\delta t=0$ (above the $T_{cs}$ ), divide the $T-\delta t$ plane into four domains with different quantum phases of the superconducting-proximity SSH chain. The non-superconducting TTP and TP appear at $\delta t > 0$ and $\delta t < 0$ above the $T_{cs}$ curve, respectively. In these two phases,  $\Delta(i)$ is equal to zero in the entire system, including the bulk and surface regions. The bulk SC phase is located below the $T_{cb}$ curve about $\delta t = 0$~(the dark green region in the figure). Finally, the domain of the surface SC phase is located between the $T_{cs}$ and $T_{cb}$ curves, where the pair potential is non-zero only near the chain edges. Here three subdomains can be classified. For $\omega_I \geq 99\%$ we have the interference-induced surface SC~(the light blue region with $\delta t\geq -0.22$). The TBS-induced surface SC corresponds to $\omega_I \leq 1\%$ (the red region with $\delta t\leq -0.9$). The intermediate (crossover) surface SC is defined by $1\% <\omega_I <99\%$ (the yellow region with $-0.9 <\delta t < -0.22$). Thus, we arrive at the main conclusion of the present work: the interference-induced surface SC plays a crucial role in the superconducting-proximity SSH chain. Surprisingly, one cannot explain the surface SC properties in STIs only by the presence of the surface bound states since the surface constructive interference of the scattering quasiparticle states is significant, as well.

\textit{Possible experimental realization.—}To experimentally observe our results, one can employ an SSH-type material such as conducting polymers~\cite{Su1979,Heeger1988}, e.g., polyacetylene and polythiophene, transition-metal monochalcogenide nanowires, e.g., M$_6$X$_6$ (M = Mo, W and X = S, Se, Te)~\cite{Jin2020,Nagata2019,Venkataraman2006,Venkataraman1999}, artificial atomic dimer chains made by vacancy sites~\cite{Drost2017}, and graphene nanoribbons~\cite{Groning2018,Rizzo2018}. The key controlling parameter ($\delta t$) for the surface SC crossover and $T_{cs}$ is associated with the energy gap ($\Delta E_{\rm min}$) between the two non-SC bands of these systems, i.e., $\delta t=\Delta E_{\rm min}/4$~\cite{Su1979,Heeger1988,Drost2017}, which can be tuned by chemical synthesis, doping, strain engineering, etc. Furthermore, according to our numerical results, it is reasonable to expect that other 1D and 2D topological insulators under the $s$-wave superconducting proximity may also exhibit interference-induced surface SC and enhanced $T_{cs}$ if the energies of the TBS-related quasiparticles are non-zero. Recall that the zero-energy TBS quasiparticles (i.e., Majorana quasiparticles) are normally hosted in the regions with a suppressed SC. 

\textit{Conclusions.—}The self-consistent numerical BdG calculations for an $s$-wave superconducting-proximity SSH chain have elucidated the emergence of the surface SC through the interference of scattering quasiparticle states, rather than due to the TBSs localized near the chain edges. The constructive interference of scattering quasiparticle states results in an enhanced surface SC near the chain edges, whereas their destructive interference tends to suppress the bulk SC in the chain center. As the system progresses deeper into the TP with decreasing the hopping staggering parameter, the surface SC undergoes a crossover from the interference-dominated to the TBS-induced regime. In this crossover, the surface critical temperature experiences a significant increase, while the bulk critical temperature remains at zero. The attainment of the maximum surface critical temperature coincides with the transition of the surface SC to the TBS-dominated regime. Our findings shed light on the nature of the surface SC in STIs and underscore the potential for experimentally modulating the superconducting critical temperature in these materials.


\begin{acknowledgments}
This work was supported by the Science Foundation of Zhejiang Sci-Tech University(ZSTU) (Grants No. 19062463-Y). The study has also been funded within the framework of the HSE University Basic Research Program.
\end{acknowledgments}


%

\end{document}